\documentclass[12pt,reqno]{amsart}
\usepackage{graphicx,color}
\usepackage[centertags]{amsmath}
\usepackage{enumerate}

\begin{document}

\title{The Problem Of Grue Isn't}

\author{William M. Briggs}
\address{New York City}
\email{matt@wmbriggs.com}
\urladdr{wmbriggs.com} 

\maketitle

The so-called problem of grue was introduced by Nelson Goodman in 1954 \cite{Goo1954} as a ``riddle" about induction, a riddle which has been widely thought to cast doubt on the validity and rationality of induction. That unnecessary doubt in turn is partly responsible for the reluctance to adopt the view that probability is part of logic: for that view see, e.g., \cite{Fra2001,Jay2003,Sto1983}. Several authors have pointed out deficiencies in grue, most notably \cite{Gro2009,Sto1983}. Nevertheless, the ``problem" still excites some authors, e.g. \cite{Sch2014} (and references therein).

Here, adapted from Groarke (p. 65), is the basis of grue, along with another simple demonstration that the ``problem" makes no sense (Groarke lists others, as does Stove).

Grue is a predicate, like green or blue, but with a built-in {\it ad hoc} time component. Objects are grue if they are {\it green} and observed before (say) 21 October 1978 or {\it fast} and observed after that date. A green grape observed 20 October 1978 and a fast (say, white) car observed 22 October 1978 are grue. But if you saw the green grape yesterday, or remember seeing the  fast car in 1976, then neither are grue. The definition changes with the arbitrary date.

Imagine it's before the Date and you've seen or heard of only green emeralds. Induction says future, or rather all unobserved, emeralds will also be green. But since it's before the Date, these emeralds are also grue, thus induction also says all unobserved emeralds will be grue. Finally comes a point after the Date, and lo, a green and not a fast emerald appears, thus not a grue emerald. Induction, which told us that emerable should be grue, is broken!

The reason we expect (via induction) unobserved emeralds to be green is we expect that whatever is {\it causing} emeralds to be green will remain the same through time. Whether this is the formal, material, efficient, or final cause depends on the perspective one takes, of course.  We comprehend the {\it essence} of what it is to be an emerald is unchanging. And that is what induction is: the understanding of this essence, an awareness of cause. Rather, that is one form of induction.  Groarke \cite{Gro2009} contends Aritotle's {\it epagoge} is analogical and there are five types of induction; see his Chapter 4 for a delineation. See David Oderberg on essentialism, \cite{Ode2008}.

Nobody has ever seen a fast emerald; neither are blithe, winsome, electrifying, salty, nor brutal emeralds observed. Nobody has ever seen a blue one either, yet it is blue that is the traditional predicate stated in the ``problem", not fast or blithe...  But the choice of predicate is arbitrary; there is nothing special about blue. Using an absurd one like fast makes the so-called problem of grue disappear, because we realize that no emerable can suddenly change nature from green to fast. That is, our understanding (via induction) that it is the essence of emeralds to be green, that some thing or things are causing the greeness, is what leads us to reject the idea that this cause can suddenly switch and create blithe or fast emeralds instead of green ones. 

Incidentally, there is no causation in the predicate grue, as has often been noted. Which is to say, the riddle does not suppose emeralds are changing their nature (meaning no change in any formal, material, efficient, or final cause takes place), but that induction is supposed to indicate that some change in nature {\it should} take place on the Date but doesn't.  After all, some thing or things must operate to cause the change.   Grue, then, is a mix up in understanding causation.

Again, we do not know of any cause (or any type) that will switch emeralds abruptly from green-mode to blue-mode. It is thus obvious that the predicate blue is what caused (in our minds) the difficulty all along. We observe that colors change in certain objects like flowers or cars. Via induction, we expect that this change is natural or is of the essence of these objects. Why? Because we're aware of the causes of color change which make the object at one time this color and at another time that color. For instance, a leaf changing from green to red on a certain date. This does not shock because we are aware of the cause of this change.  Amusingly, if we re-create the grue ``problem" for the leaf using green and red, and we get the right date, then grue-type induction {\it works} for autumn leaves.

There was never anything wrong with induction. Far from causing us to doubt induction, thinking about grue strengthens the confidence we have in it because we realize that grue seemed problematic because it tortured our understanding of what caused emeralds to be green.

\bibliographystyle{plain}
\bibliography{/home/briggs/projects/writing/logic}

\begin{thebibliography}{1}

\bibitem{Fra2001}
James Franklin.
\newblock Resurrecting logical probability.
\newblock {\em Erkenntnis}, 55:277--305, 2001.

\bibitem{Goo1954}
Nelson Goodman.
\newblock {\em Fact, fiction, and forecast}.
\newblock Bobbs-Merrill, Indianapolis, 1954.

\bibitem{Gro2009}
Louis Groarke.
\newblock {\em An Aristotelian Account of Induction}.
\newblock Mcgill Queens University Press, Montreal, 2009.

\bibitem{Jay2003}
E.~T. Jaynes.
\newblock {\em Probability Theory: The Logic of Science}.
\newblock Cambridge University Press, Cambridge, 2003.

\bibitem{Ode2008}
David Oderberg.
\newblock {\em Real Essentialism}.
\newblock Routledge, London, 2008.

\bibitem{Sch2014}
Alfred Schramm.
\newblock Evidence, hypothesis, and grue.
\newblock {\em Erkenntnis}, 79(3):571--591, 2014.

\bibitem{Sto1983}
David Stove.
\newblock {\em The Rationality of Induction}.
\newblock Clarendon, Oxford, 1986.

\end{thebibliography}
%\bibliography{logic}

\end{document}